\documentclass{article}
\usepackage{graphics}
\usepackage{latexsym}

\begin{document}

\begin{titlepage}
\begin{flushright}
LU TP 99-32 \\
October 1999
\end{flushright}
\vspace{25mm}
\begin{center}
\Large
{\bf The Diagonalisation of the Lund Fragmentation Model I} \\
\normalsize
\vspace{12mm}
Bo Andersson\footnote{bo@thep.lu.se} \vspace{1ex} \\
Fredrik S\"oderberg\footnote{fredrik@thep.lu.se} \vspace{1ex}\\
Department of Theoretical Physics, Lund University, \\
S\"olvegatan 14A, S-223 62 Lund, Sweden 
\end{center}
\vspace{20mm}
\noindent {\bf Abstract} \\
We will in this note show that it is possible to diagonalise the Lund Fragmentation Model. We show that the basic original result, the Lund Area law, can be factorised into a product of transition operators, each describing the production of a single particle and the two adjacent breakup points (vertex positions) of the string field. The transition operator has a discrete spectrum of (orthonormal) eigenfunctions, describing the vertex positions (which in a dual way corresponds to the momentum transfers between the produced particles) and discrete eigenvalues, which only depend upon the particle produced. The eigenfunctions turn out to be the well-known two-dimensional harmonic oscillator functions and the eigenvalues are the analytic continuations of these functions to time-like values (corresponding to the particle mass). In this way all observables in the model can be expressed in terms of analytical formulas. In this note only the $1+1$-dimensional version of the model is treated but we end with remarks on the extensions to gluonic radiation, transverse momentum generation etc, to be performed in future papers. 
\end{titlepage}

\section{Introduction}
\label{intro}

The Lund Fragmentation Model is built upon a few very general assumptions: there is a string-like force field between the coloured constituents, there is causality and Lorentz covariance, the production of the particles can be described in terms of a stochastical process and the process will obey a saturation hypothesis. Using semi-classical probability considerations we are then led, \cite{BS},\cite{BA}, to a unique stochastical process for the breakup of the force field into the final state hadrons. In Section \ref{fragmentation}, we will provide a set of necessary formulas to describe the dynamical developments along the surface, spanned by the string field. The major result is that the probability to reach a particular (exclusive) final state is given by the phase space of the state multiplied by a negative exponential of the area spanned before the string decays (``the Lund area-law''). In general the model has been used as it is implemented in the well-known Monte Carlo simulation program \textsc{Jetset}, \cite{TS}. This means, on the one hand, that it is possible to take into account a large amount of kinematical complications, in particular from the decay of the primary produced resonances. On the other hand, in order to make the simulation programs time-effective, it is necessary to introduce routines that make the process rather difficult to follow. In particular, it is difficult to disentangle the major dynamical features of the model from the many necessary numerical compromises in the simulation program. 

In this note, we will show how to diagonalise the basic stochastical process, i.e. how to define a complete set of eigenfunctions and eigenvalues describing the process on the local level. In this way we can provide analytical formulas for all possible correlations between the observables in the process. 

In Section \ref{transition}, we will factorise the Lund area law in a somewhat different way than the ordinary. It is done by defining an operator that describes the transition from one production point to the next in the process. It can be written as an integral operator describing the probability to go from one breakup point along the string field to the next thereby producing a particular hadron. We show how to diagonalise the operator in terms of its eigenfunctions (which in a very neat way corresponds to two-dimensional harmonic oscillator functions) and calculate its eigenvalues, that are closely related to the Lund fragmentation function. 

These eigenvalues turn out to be the major building stones in all the model correlations and in Section \ref{sdependence}, we show a set of properties of the eigenvalues. It turns out that the eigenvalues form a discrete set, and that they
correspond to analytical continuations of the harmonic oscillator eigenfunctions from space-like to time-like regions. We show how to use the factorisation properties of the model to provide a set of useful relations for the products of the eigenvalues. We also exhibit the relationship to a field theory in a two-dimensional Euclidian space that will be further pursued in future publications.

We will in this note be satisfied to treat only the simplest case corresponding
to the $1+1$-dimensional dynamics of the Lund Model. We will, however, end with an outlook on future work, in particular on the effects of gluon emission and transverse momentum properties of the hadronisation process.
\section{The Lund Fragmentation Model}
\label{fragmentation}

\begin{figure}[h]
\begin{center}
\includegraphics{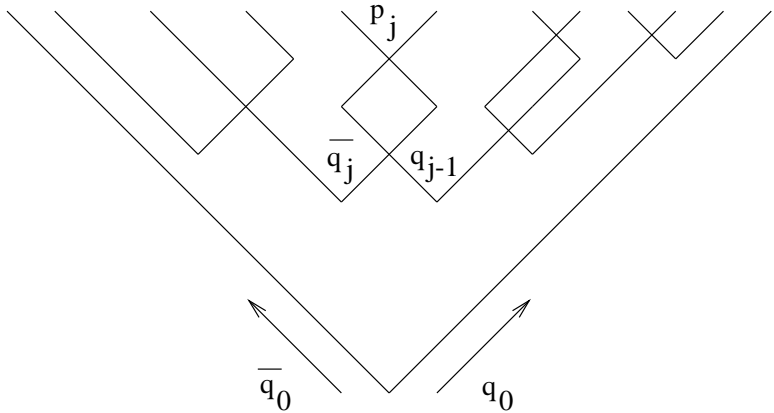}
\end{center}
\caption{A high-energy string breakup of an original $q_{0} \overline{q_{0}}$ -pair, created at a single space-time point.} \label{spacetime}
\end{figure}
The massless relativistic string is in the Lund Model used as a model for the colour force fields with colour-$3$ quarks ($q$) and colour-$\bar{3}$ antiquarks ($\bar{q}$) at the endpoints. Finally, the colour-$8$ gluons ($g$) are in the Lund Model interpreted as internal excitations on the string. We will, in this note, treat all the $(q\bar{q})$-particles as massless and moving along the lightcones. However, the final result is independent of this assumption, cf. \cite{BA}, (massive $q$ and $\bar{q}$ would in a semi-classical scenario move along hyperbolas with the lightcones as asymptotes). Further, we will as examples of the formalism consider only $e^+e^-$-annihilation reactions and refrain other processes as well as gluonic bremsstrahlung to future work. 

Then an original $q_o \bar{q}_o$-pair is assumed to be created at a single space-time point and start to go apart thereby stretching the string field in between them (cf. Fig \ref{spacetime}). The field will break up into new pairs at the vertices $x_{j}=(x_{j +}, x_{j -})$ (we use lightcone coordinates) and a $q\equiv q_{j-1}$ will together with a  $\bar{q} \equiv \bar{q}_{j}$ from the adjacent vertex form a final state hadron with the energy-momentum $p_j$. 

In this way we have introduced a convenient ordering in the form of rank: the first rank particle is formed by $(q_o \bar{q}_1)$, the second rank by $(q_1
\bar{q}_2)$ etc. It is also possible to introduce a rank-ordering from the
$\bar{q}_o$ side, i.e. along the opposite lightcone starting with $(\bar{q}_o q_{n-1})$ in an $n$-particle final state. The dynamical results should of course be independent of the ordering. Actually, it is easy to convince oneself that it is necessary that all the vertices must be space-like with respect to each other. One finds that the energy-momentum of the $j$-th particle is given by $p_j=\kappa(x_{j-1+}-x_{j+},x_{j-}-x_{j-1-})$ (here $\kappa \simeq 1$ $GeV/fm$ is the string constant and we will for simplicity put it equal to unity). As the vector $p_j$ must be time-like (with squared mass equal to $m^2_j=p_j^2$) we conclude that the two adjacent vertices are space-like.

\begin{figure}
\begin{center}
\includegraphics{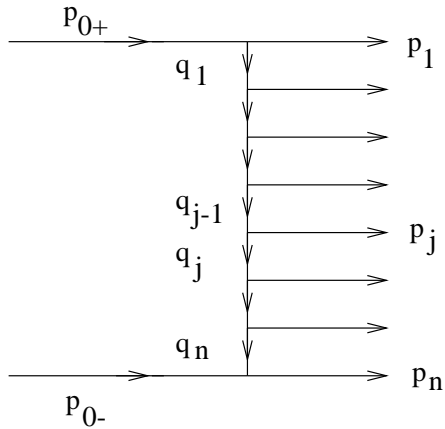}
\end{center}
\caption{The production process described in terms of momentum transfers in a chain.} \label{qLund}
\end{figure}

There is immediately a second conclusion. If we define the vectors $q_j=(x_{j+},-x_{j-})$ then we obtain that $p_j=q_{j-1}-q_{j}$. This means that the Lund fragmentation process also can be described by means of a ladder graph as in Fig. \ref{qLund}. Thus the energy-momentum transfers between the particles along the ladder is in a dual relationship to the production vertices in the description in Fig. \ref{spacetime}. We will use this relationship in Section \ref{transition} to derive the results of this paper.

The derivation of the Lund Model formulas is based upon these observations and
a final assumption: even if the energy of the original pair (in the cms denoted
 conventionally $W=\sqrt{s}$) will increase without limit the distribution of the proper times of the production vertices $\Gamma_j=x_j^2= x_{j+}x_{j-}$ will stay finite. In terms of the momentum transfers this ``saturation assumption'' evidently corresponds to (one of) the ordinary assumptions behind Gribov's Reggeon theory, that the momentum transfers stay finite in this limit.

\begin{figure}
\begin{center}
\includegraphics{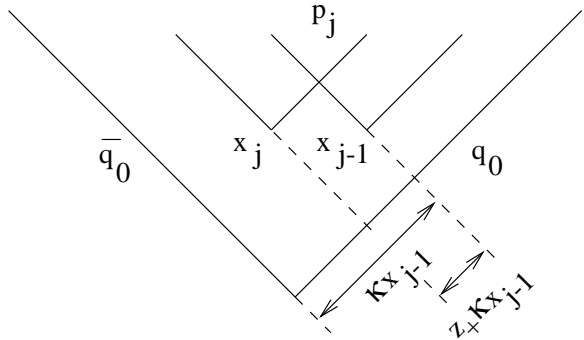}
\end{center}
\caption{Two adjacent vertices with coordinates $x_{j}$ and $x_{j-1}$ (in the center of the process) and a hadron with energy-momentum $p_{j}$ produced in between.} \label{centre}
\end{figure}

To see the details, we will concentrate on two adjacent vertices in the center of the process with the coordinates $x_j$ and $x_{j-1}$ such that the above-mentioned hadron with $p_j$ is produced in between, cf. Fig. \ref{centre}. It is convenient to introduce the coordinates $z_+=1-x_{j+}/x_{j-1 +}$ and $z_-=1-x_{j-1 -}/x_{j-}$ that are Lorentz invariants and will have the range $0\leq z_{\pm} \leq 1$ 
independent of the other variables. We also describe the vertices by the hyperbolic coordinates $(\Gamma_{\ell},y_{\ell})$, $\ell=j-1,j$ and note that due to Lorentz covariance the process can only depend upon the $\Gamma$'s (i.e. the proper times squared) and the relative hyperbolic angles $\delta y_j= y_{j-1}-y_j$ (note that the $\delta y$'s will be fixed by the mass requirements). 

We will then consider the breakup vertex $x_{j-1}$ to be the last in a long row of production points along the positive lightcone and using the saturation assumption we expect that the probability distribution is
\begin{eqnarray}
\label{Hdef}
H(x_{j-1}) dx_{j-1 +}dx_{j-1 -} \equiv
H(\Gamma_{j-1})d\Gamma_{j-1}dy_{j-1}
\end{eqnarray}
In the second line we have made use of Lorentz invariance to claim that the function $H$ only depends upon $\Gamma_{j-1}$. The production of the particle $p_j$ is then given by another ``step'' along the positive lightcone with the probability $f(z_+) dz_+$ to take the fraction $z_+$ of the remainder. Then the combined probability is given by
\begin{eqnarray}
\label{j-1def}
H(\Gamma_{j-1})d\Gamma_{j-1}dy_{j-1} f(z_+) dz_+
\end{eqnarray}
In the same way we may consider the production of the particle as the last in a
long row of steps along the negative lightcone, firstly arriving at the vertex $x_j$ with the probability $H$ and then taking another step along the negative lightcone. In this way we obtain the joint probability
\begin{eqnarray}
\label{jdef}
H(\Gamma_{j})d\Gamma_{j}dy_{j} f(z_-) dz_-
\end{eqnarray}
The basic assumption in the Lund Model is then that the two probability distributions in Eqs. (\ref{j-1def}) and (\ref{jdef}) are equal and this defines in a unique way the distributions $H$ and $f$,\cite{BS},\cite{BA}
\begin{eqnarray}
\label{fHdef}
& &H_j(\Gamma) = C_j \Gamma^{a_j} \exp(-b\Gamma) \nonumber\\
& &f_{j-1,j}(z)= N_{j-1,j} (1-z)^{a_j} z^{a_{j-1}-a_{j}-1} \exp(-bm^2/z)
\end{eqnarray}
The parameters $a_j$ (with the notation for $f_ {j-1,j}$ meaning that the hadron with mass $m$ is produced in a step from the point $j-1$ to the point $j$) may be different for different vertices (e.g. spin- and/or flavour dependent) but the parameter $b$ should be the same, i.e. it must correspond to a general colour dynamical property. (Speculations on its origin can be found in e.g. \cite{BA}). In the phenomenological applications of the Lund Model there has (besides the first particle in a heavy quark jet according to a suggestion by Bowler, \cite{Bow}) been no use for more than a common $a$-value. We will in general in this paper treat this simpler case and only when it is useful exhibit the differences to the general case. Finally, the parameters $C_j$ and $N_{j-1,j}$ are normalisation constants.

\begin{figure}
\begin{center}
\includegraphics{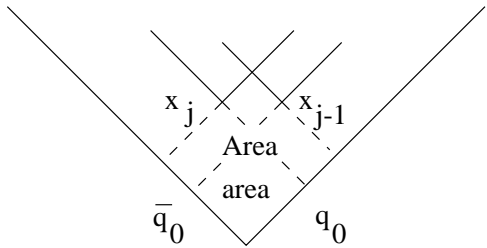}
\end{center}
\caption{The two areas used in Eq. (\ref{fHtog}). The large area is the one spanned below the first meeting point of the two constituents from the adjacent vertices.} \label{Aarea}
\end{figure}

The joint probability distribution $H(\Gamma)f(z)$ can then be written as
\begin{eqnarray}
\label{fHtog}
H(\Gamma)f(z)\propto C N[(1-z)\Gamma]^a \exp(-b(\Gamma+m^2/z) \equiv
(area)^a
\exp(-b (Area))
\end{eqnarray}
where the two areas, the large and the small one, are shown in Fig. \ref{Aarea}. Evidently, the opposite production direction will produce the same result and the areas play therefore due their simple factorisation properties (just as they do in general for gauge field theories) a fundamental role in the process. (If there are different values of the $a$-parameter the result is similar with different areas represented, one typical of each vertex).

\begin{figure}
\begin{center}
\includegraphics{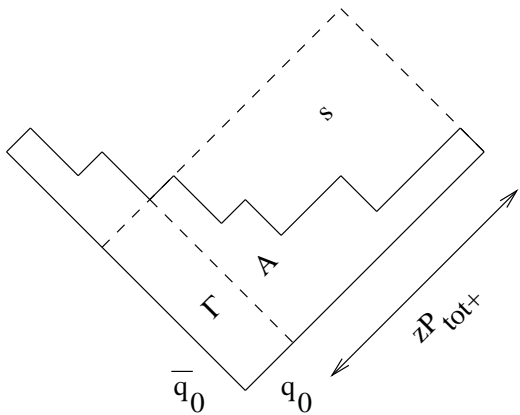}
\end{center}
\caption{An N-particle cluster with notation as explained in the text.} \label{nparticle}
\end{figure}

We will next consider the probability to produce a rank-connected $N$-particle set $\{p_j\}$, for definiteness along the positive lightcone starting at the turning point of the original $q_o$ (cf. Fig. \ref{nparticle}). The first rank particle will then take a fraction $z_1$ of the total lightcone energy momentum, the second will take a fraction $z_2$ of what is left, i.e. $(1-z_1)$ etc. The observable fractions are then $\zeta_1\equiv z_1$, $\zeta_2 =z_2(1-\zeta_1)$, $\zeta_3=z_3(1-\zeta_1-\zeta_2)=z_3(1-z_1)(1-z_2)$ etc. In this way we obtain in easily understood notations, \cite{BS},\cite{BA}:
\begin{eqnarray}
\label{ncluster}
& &\prod_1^N N_j \frac{dz_j}{z_j} (1-z_j)^a \exp(-bm_j^2/z_j) =\nonumber
\\
& &\prod_1^N N_j \frac{d\zeta_j}{\zeta_j} (1-\sum_1^N \zeta_j)^a
\exp(-b(A+\Gamma))= \nonumber\\
& &ds \frac{dz}{z}(1-z)^a\exp(-bs(1-z)/z) \prod_1^N N_j 
\frac{du_j}{u_j} \times \nonumber\\
& &\exp(-bA)
\delta(1-\sum_1^N u_j) \delta(s-\sum_1^N m_j^2/u_j) \equiv dP_{ext} dP_{int}
\end{eqnarray}
In the second line, we have introduced the variables $\zeta_j$ defined above and in the third the common variable $z=\sum_1^N \zeta_j$ and finally rescaled the fractions into $u_j=\zeta_j/z$. We have also introduced the total area (according to Fig. \ref{nparticle}) $A_{tot}=A +\Gamma$ and the total cms energy $\sqrt{s}$ of the $N$-particle cluster. We note that the ``final'' vertex proper time squared is $\Gamma=(1-z)s/z$. 

The final result is then that the probability distribution can be factorised into two parts. One of them ($dP_{ext}$) (note that it is independent of the multiplicity $N$) is the probability to make a cluster of mass $\sqrt{s}$ and lightcone fraction $z$. We note the close similarity to the fragmentation function for a single particle $f$ in Eq.(\ref{fHdef}). The other one ($dP_{int}$) is  the probability that the cluster will decay into just these particular particles with the fractional energy momenta $u_j$ in the cluster cms.

The distribution $dP_{ext}$ for the general case when several values of $a$ occurs will only depend upon the first and the last $a$-values: 
\begin{eqnarray}
\label{generalcase}
dP_{ext}=ds \frac{dz}{z} z^{a_{0}}( \frac{1-z}{z})^{a_{n}} exp(-b\Gamma)
\end{eqnarray}

The distribution $dP_{ext}$ can be used to study the convergence of the ``saturation assumption'' on the distribution $H$. One finds, \cite{BA} that for squared cms-energies $s$ larger than a few times the inverse of the parameter $b$ there is an exponential convergence.

The distribution $dP_{int}$ can be reformulated using that $du_j/u_j$ is equivalent to $d^2 p \delta(p^2-m^2)$ and that the two delta functions in the same way can be written in terms of the particle energy momenta ($p_j = (u_jP_{tot +}, m^2/u_jP_{tot +})$), $P_{tot}=(P_{tot+},s/P_{tot+}$)
\begin{eqnarray}
\label{energymomcons}
\delta(1-\sum_1^n u_j) \delta(s-\sum_1^n m_j^2/u_j)= \delta^2(\sum_1^n
p_j-P_{tot})
\end{eqnarray}
Putting this together we obtain the Lund Model area law:
\begin{eqnarray}
\label{arealaw}
dP_{int} = \prod_1^n N_j d^2p_j \delta(p_j^2-m_j^2) \delta (\sum_1^n
p_j-P_{tot})
\exp(-bA)
\end{eqnarray}
In the case of different $a$-parameter values we obtain an extra factor for each $j$: $u_j^{(\delta a)_j}$ with $(\delta a)_j =a_{j-1}-a_{j}$, i.e. only the differences of the adjacent $a$-values occur. (It is useful to note that $u_{+j}u_{-j}=m_j^2/s$ in order to see that the formula is symmetric between the forward and backward lightcones).

\section{The Transition Operator and its Eigenfunctions}
\label{transition}
We will in this section rearrange the Lund Model area law , cf. Eq. (\ref{arealaw}), in another form i.e. as a product of a set of step operators taking us from one vertex to the next thereby producing a particle in between. We will after that show that this transition operator has a well-defined set of eigenfunctions with discrete eigenvalues.

\begin{figure}[h]
\begin{center}
\includegraphics{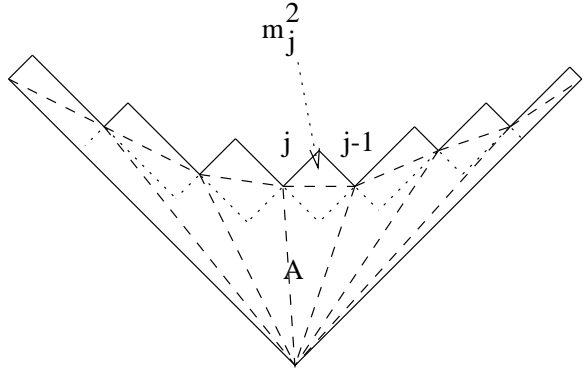}
\end{center}
\caption{The area A subdivided into triangular regions as described in the text.} \label{step}
\end{figure}

To do that, we note that the area $A$ can according to Fig. \ref{step} be subdivided into (hyperbolic) triangular regions (each with an extra ``tip'' corresponding to half of the squared mass; we neglect them for the moment and note that they can be included in the particle production constants $N_j$). The size of these regions are given by $(\delta A)_j = \sqrt{\Gamma_{j-1} \Gamma_j} |sinh (\delta y_j)|$ (we are using the notations from the earlier section) and one
finds by direct calculation that
\begin{eqnarray}
\label{deltaA}
& &(\delta A)_j = \sqrt{\lambda(\Gamma_{j-1},\Gamma_j,-m_j^2)}/2
\nonumber \\
& &\lambda(a,b,c)=a^2+b^2+c^2-2ab-2bc-2ac
\end{eqnarray}
If we introduce the lightcone fraction $z$ according to Fig. \ref{step}, then we
can write
\begin{eqnarray}
\label{zdefine}
& &\Gamma_{j-1}=(1-z)(\Gamma_j+m^2_j/z) \nonumber\\ 
& &(\delta A)_j=(z \Gamma_j +m^2_j/z)/2
\end{eqnarray}
We also note the vectors $q_j=(x_{j+},-x_{j-})$ (with $-q_j^2= \Gamma_j$) that fulfil $p_j=q_{j-1}-q_{j}$, thereby ``solving'' the energy-momentum conservation conditions in Eq.(\ref{arealaw}). If we introduce them instead of the particle
vectors $\{p_j\}$, then we find the Jacobian 
\begin{eqnarray}
\label{Jacobian}
d^2p_j \delta(p_j^2-m_j^2)=d\Gamma_j
/\sqrt{\lambda(\Gamma_{j-1},\Gamma_j,-m_j^2)} 
\end{eqnarray}
We can consequently subdivide the whole process according to the arealaw as
\begin{eqnarray}
\label{Kdef}
& &dP_{int} = \prod_1^n K(\Gamma_{j-1},\Gamma_j,m_j^2) d\Gamma_j
\nonumber \\
& &K(\Gamma_{j-1},\Gamma_j,m_j^2)=N_j
\frac{ \exp(-b/2 \sqrt{ \lambda( \Gamma_{j-1},\Gamma_j,-m_j^2)})}{\sqrt{\lambda(\Gamma_{j-1},\Gamma_j,-m_j^2)}}
\end{eqnarray}
A useful representation of the kernel function $K$ is (it is easily obtained from the considerations above)
\begin{eqnarray}
\label{Krepresent}
& &K( \Gamma_{j-1},\Gamma_j,m_j^2)= \nonumber \\ 
& &\int_0^1 \frac{dz \exp(-b(z \Gamma_j+m^2_j/z)/2)}{z}
\delta(\Gamma_{j-1}-(1-z)(\Gamma_j+m^2_j/z))
\end{eqnarray}
(To be precise the result in Eq.(\ref{Kdef}) must be supplemented by boundary conditions but we will neglect them because in this paper we will only be interested in results outside the fragmentation regions). It is useful to consider the eigenfunctions of the transition operator $K$. For simplicity we will introduce the dimensionless variables  $\Gamma_{\ell} \rightarrow b\Gamma_{\ell}\equiv x_{\ell}$ and $m^2_{\ell} \rightarrow bm^2_{\ell} =y_{\ell}$ and consider the solution to the equations
\begin{eqnarray}
\label{eigendef}
\lambda_n g_n(x) = \int K(x,x^{\prime},y)g_n(x^{\prime})dx^{\prime}
\end{eqnarray}
The surprising and very gratifying result we obtain is that the functions $g_n$ are well-known in mathematical analysis. We are going to call them the Laguerre functions (noting that the Laguerre polynomials $L_n$ are orthonormal in the measure $dx \exp(-x)$)
\begin{eqnarray}
\label{gdef}
g_n(x)=L_n(x) \exp(-x/2)
\end{eqnarray}
They are orthonormal on the positive real axis $0\leq x\leq \infty$ in the measure $dx$ (we use the notations from \cite{MagnusOber}) 
\begin{eqnarray}
\label{ortho}
& &\int_0^{\infty} dx g_n(x)g_m(x)=\delta_{n,m} \nonumber\\
& &\sum_n g_n(x)g_n(x^{\prime}) = \delta (x-x^{\prime}) 
\end{eqnarray}
Further, the Laguerre functions are the eigenfunctions of the two-dimensional harmonic oscillator corresponding to angular momentum equal to zero. In fact, it is easy to prove that the eigenfunctions $g_n$ will fulfil the following equation because of the well-known properties of the Laguerre polynomials $L_n$  
\begin{eqnarray}
\label{diffeqn}
& &x\frac{d^2L_n}{dx^2}+(1-x)\frac{dL_n}{dx}+nL_n=0 \nonumber\\
& &(-\bigtriangleup + bQ^2)g_n(bQ^2)=2(2n+1)g_n(bQ^2)
\end{eqnarray}
In the second line, we have considered the two-component vector $Q$ with the scalar product $Q^2=Q_1^2\pm Q_0^2$. The differential operator is correspondingly defined as $\bigtriangleup= \frac{\partial^2}{\partial bQ_1^2} \pm \frac{\partial^2}{\partial bQ_0^2}$ i.e. the equation is valid both for Euclidian metric and for space-like directions in two-dimensional Minkowski space.

We note in particular, that the eigenfunctions are independent of the mass $m$ of the particle produced in between, this mass-dependence comes solely in the eigenvalues
\begin{eqnarray}
\label{lambdadef}
\lambda_n(y)=N \exp(y/2) \int_0^1 dz/z \exp(-y/z) (1-z)^n
\end{eqnarray}
Here we have kept to the definition of the hyperbolic triangle, without the ``tip'', in the kernel function $K$  (cf. the remarks before the Eq.(\ref{deltaA})). We will be concerned with the properties of the eigenvalues in the next section but we note at this point their close relationship to the Lund fragmentation function $f$ (for equal $a$-values) in Eq.(\ref{fHdef}). It is also obvious from
Eq.(\ref{lambdadef}) that the eigenvalues will be discrete and decrease quickly with $n$.

To obtain these results from Eq.(\ref{eigendef}), we make use of the representation of the kernel function $K$ in Eq.(\ref{Krepresent}) and find the following necessary (and sufficient) requirements on the Laguerre polynomials:
\begin{eqnarray}
\label{mastereq}
\lambda_n L_n(x) = \int_0^1 dz/z \exp(-y/z) L_n((1-z)(x+y/z))
\end{eqnarray}
It is easy to see that for a polynomial of the $n$:th degree the eigenvalues $\lambda_n$ will have to fulfil Eq.(\ref{lambdadef}). However, to prove the general result in Eq.(\ref{mastereq}) we have expanded both sides in powers, performed for the $m$:th term inside the integral $0 \leq \ell \leq m$ partial integrations (to get rid of the powers of $y$) and then gathered the powers in $x$. We feel that there must be a simpler way but we have not found it yet. 

\begin{figure}
\begin{center}
\includegraphics{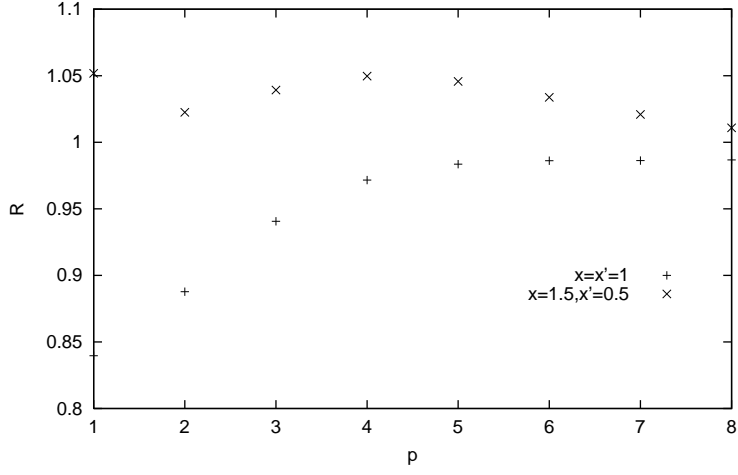}
\end{center}
\caption{The ratio R=$\frac{ \exp(- \frac{1}{2} \sqrt{ \lambda(x,x^{\prime},-y)})}{ \sqrt{ \lambda(x,x^{\prime},-y)}}/ \sum_{n=0}^{p} g_n(x) \lambda_n(y) g_n(x^{\prime})$ for different values of p,x and $x^{\prime}$ when y=0.5.} 
\label{diagram}
\end{figure}

Given these results, we note that a theorem attributed to Mercer (private information, \cite{EdW}) provides the following representation for the transition operator $K$:
\begin{eqnarray}
\label{representK}
K(x,x^{\prime},y)= \sum_{n=0}^{\infty} g_n(x) \lambda_n(y)
g_n(x^{\prime})
\end{eqnarray}
In order to check the convergence properties of Eq.(\ref{representK}), we show in Fig. \ref{diagram} the results for the ratio of the left-hand to the right-hand side of the equation. It is evidently in general only necessary to keep a few terms to obtain a good approximation.

Due to the orthonormality of the Laguerre functions, it is further immediately obvious that while Eq.(\ref{representK}) represents the distribution after a single particle production between $x^{\prime}$ and $x$ the result for the production of $N$ particles in between them is given by
\begin{eqnarray}
\label{NpartK}
K_N(x,x^{\prime}) = \sum_{n=0}^{\infty} g_n(x) (\lambda_n(y))^N
g_n(x^{\prime})
\end{eqnarray}
In the next section, we will show how to provide a formula for a fixed invariant mass-square $s$ and/or a fixed lightcone fraction $z$ (thereby completely defining the relationships) between the points labelled by $x^{\prime}$ and $x$.

\section{The Properties of the Eigenvalues}
\label{sdependence}

We will in the following discussion for simplicity put the normalisation constant $N$ equal to unity but we will insert it in the end formulas. Then the eigenvalues $\lambda_n(y)$ (with $y=bm^2$) defined in Eq.(\ref{lambdadef}) will have the following property:
\begin{eqnarray}
\label{lambdaprop1}
& &\exp(y/2) \lambda_n(y)= \int_0^1 dz (1-z)^n \frac{\exp(-(1-z)y/z)}{z}
= \nonumber \\
& & \int_0^1 dz \sum_m (1-z)^{m+n} L_m(y)=\sum_m \frac{L_m(y)}{m+n+1}
\end{eqnarray}
We have here in going from the second to the third line in Eq. (\ref{lambdaprop1}) made use of  the generating function for the Laguerre polynomials (\cite{MagnusOber}):
\begin{eqnarray}
\label{genfunct}
\sum_n L_n(y)z^n = \sum_k \frac{(-y)^k}{k!} \sum_m \frac{z^{m+k}
(m+k)!}{m!
k!}= \frac{\exp[-yz/(1-z)]}{(1-z)}
\end{eqnarray}
In the second line, we have introduced the series expansion for the Laguerre polynomials, rearranged it by changing the original index $n\rightarrow m+k$ and then summed up firstly a negative binomial and then an exponential series.  

Consequently, the eigenvalue for $n$ can be written as a series in the
eigenfunctions $g_m$ and we may from this representation immediately
conclude
(using the differential equation in Eq.(\ref{diffeqn})) that the
eigenvalues will fulfil
\begin{eqnarray}
\label{diffeqn2}
(-\bigtriangleup + bp^2)\lambda_n(bp^2)=-2(2n+1)\lambda_n(bp^2) +4
\delta(bp^2)
\end{eqnarray}
Here we have introduced the timelike vector $p$ with length equal to the mass (and the differential operator $\bigtriangleup$ is defined in terms of $p$:s components). We have also used the result in Eq.(\ref{ortho}) noting that $g_m(0)=1$ for all values of $m$. We conclude that also the eigenvalues are governed by the harmonic oscillator equation and that they correspond to a particular analytic continuation of these functions from spacelike to timelike vectors. 

Actually, the eigenvalues $\lambda_n(y)$ are solutions to the (degenerate) hypergeometric differential equation and in conventional notations, \cite{MagnusOber}  in terms of Whittaker functions we have $\lambda_n(y) = n! W_{-n-1/2,0}(y)/\sqrt{y}$. Using either the differential equation in Eq.(\ref{diffeqn2}) or the formulas in \cite{MagnusOber} we obtain another useful representation that bears out
these analyticity properties:
\begin{eqnarray}
\label{lambdaagain}
\lambda_n(y) = \int_0^{\infty} dt \frac{g_n(t) \exp(-(t+y)/2)}{t+y}
\end{eqnarray}
Next we will consider the correlation coefficients for the case when we produce $N$ particles in between the vertices denoted $x$ and $x^{\prime}$ in Eq. (\ref{NpartK}). Using the same procedure as in connection with the derivation of the $N$-particle cluster in Eq.(\ref{ncluster}), we can immediately write
\begin{eqnarray}
\label{Ncluster}
& &\lambda_n^N(y)= \int ds R_N(s) \hat{\lambda}_n(bs)\nonumber\\
& &R_N(s) = \int \prod_1^N N_j d^2p_j \delta(p_j^2-m_j^2) 
\delta (\sum_1^N p_j-P_{tot}) \exp(-bA)\nonumber\\
& & \hat{\lambda}_n(bs)= \exp(bs/2) \lambda_n(bs)
\end{eqnarray} 
The quantity $\hat{\lambda}_n(bs)$ is the probability (in the $n$th harmonic
oscillator state) to produce a cluster with the energy $s$. It is the integral over all $z$ values of $dP_{ext}$ in Eq. (\ref{ncluster}) (with the Lund parameter $a$ exchanged for $n$). In the same way $R_N(s)$ is related to the integral over $dP_{int}$ in Eq. (\ref{ncluster}), i.e. it is the phase space integral (including the area law) of the $N$ particles. We have brought back the normalisation constants $N_j$ in the expression for $R_N$.

It is now evident how to obtain the distributions of $x$ and $x^{\prime}$ when
there are $N$ particles with a fixed squared mass $s$ in between:
\begin{eqnarray}
\label{sNcorr}
\sum_n g_n(x) \hat{\lambda}_n(bs) R_N(s) g_n(x^{\prime})
\end{eqnarray}
We will next turn to the properties of the phase space integrals $R_N$ but before that we make the following observation. The model has very  simple factorisation properties both in  the energy-momentum fractions and, as we have shown in this note, in the energy-momentum transfers (or in the dual language, in 
the vertex positions). Consequently, it is in the same way as  in Eq.(\ref{sNcorr})  possible to pick out any other variables, like particular energy-momentum fractions somewhere ``in between'' and reformulate the remaining correlation coefficients accordingly. (It is of course necessary to define the scaling variables properly). As every possible observable is either related to the energy momentum transfers or to the energy-momentum of the observed particles we have in this way a complete analytical description of the process.

We will show two particularly simple and useful properties of the phase space integrals $R_N$. They fulfil a set of iterative integral equations and there are very simple formulas for the analytic function ${\cal R}_N$ which is defined by 
\begin{eqnarray}
\label{calR}
{\cal R}_N(u) = \int ds \frac{R_N(s)}{s+u}
\end{eqnarray}
To see these properties we note that if we ``pick out'' the first rank particle
from the $N$-particle cluster that defines $R_N$ we obtain, cf. Eqs. (\ref{ncluster}) and (\ref{Ncluster}) $N_1du_1/u_1 \exp(-bm^2/u_1)$.  The remaining $(N-1)$ particles  will give the same contribution but the energy is reduced to
$s_1=(P_{tot}-p_1)^2=(1-u_1)(s-m^2/u_1)$. We have consequently the integral equation
\begin{eqnarray}
\label{RNitera}
R_N(s)=\int_0^1\frac{N_1 du_1}{u_1} R_{N-1}((1-u_1)(s-m^2/u_1))
\exp(-bm^2/u_1)
\end{eqnarray}
We note the similarity to the original integral equations for the harmonic oscillator functions $g_n(x)$, but there are two major differences. The first is
the change of sign in front of the term $m^2/u_1$ in the integral (reflecting the fact that we are going from the space-like vectors $q_j$ to the timelike
vector $P_{tot}$). The second difference is that the argument in the functions
$R_N$ and $R_{N-1}$ do not have the same range. It is evident that the threshold for producing $N$ particles is $s_{N,thresh}=N^2m^2$ which is larger than the threshold for $N-1$ particles.

There is, however, another relation which will make it possible to calculate the functions $R_N$ analytically at least as a perturbation series. In order to see 
that, we consider the following sum and use the results from Eq.(\ref{Ncluster}):
\begin{eqnarray}
\label{Rcaldef}
& & \sum_n (\lambda_n(y))^N L_N(bu)=\nonumber\\
& & \int ds R_N(s) \int \frac{dz}{z} 
\exp-bs(1-z)/z \sum_n (1-z)^n L_n(bu)= \nonumber\\
& & \int ds R_N(s)\int \frac{dz}{z^2} \exp-(bs +bu)(1-z)/z = {\cal
R}_{N}(u)/b 
\end{eqnarray}
We have once again used the generating function in Eq.(\ref{genfunct}) and performed the $z$ integral. In this way we have a representation of the analytic function ${\cal R}$ along the positive real $u$-axis. It is , however, necessary to extend this function to the negative real axis in order to obtain the properties of $R_N$. We note in passing that the first line in Eq.(\ref{Rcaldef}) is (besides a factor $\exp(-bu/2)$  the contribution for the case when we would start out at the lightcone $x=0$ and consider $N$ steps to the point $x^{\prime}=bu$ ($g_n(0)=1$ for values of $n$). This result can be used in a perturbation theory by noting that for large values of $n$ we have the following behaviour of our functions: 
\begin{eqnarray}
\label{asymptot}
g_n(x) \simeq J_0(2 \sqrt{n x}) & \mbox{and} & \lambda_n(y) \simeq
\sqrt{2/\pi}
K_0(2\sqrt{ny})
\end{eqnarray}
We can then in this approximation write 
\begin{eqnarray}
\label{vdef}
& &\sum_n g_n \rightarrow \int \frac{d^2 v}{2\pi}
\exp(i2\vec{v}\vec{\mu})
\nonumber \\
& & \lambda_n(y) \rightarrow \int \frac{d^2 t}{(2\pi)^{3/2}}
\frac{\exp(i2\vec{v}\vec{t})}{\vec{t}^2 +y}
\end{eqnarray}
We have then defined two-dimensional Euclidian vectors with $\vec{\mu}^2=u$ and with $\vec{v}^2 \rightarrow n$. Therefore the whole expression can approximately be written
\begin{eqnarray}
\label{master2}
& &\exp(-bu/2) {\cal R}(u) = \sum_n (\lambda_n(y))^N g_n(bu) 
\rightarrow \nonumber \\
& &\int \prod_{j=1}^N 
\frac{d^2t_j}{(2\pi)^{3/2}(\vec{t}_j^2 +y)}\int \frac{d^2v
\exp(i2\vec{v}(\vec{\mu}-\sum\vec{t}_j))}{2\pi} = \nonumber \\
& &\int \prod_{j=1}^N 
\frac{d^2t_j}{(2\pi)^{3/2}(\vec{t}_j^2 +y)} \delta(\vec{\mu}-\sum
\vec{t}_j)/4
\end{eqnarray}
In this way we have exhibited the analytical function $\exp(-bu/2){\cal R}_N(u)$ for values of $u \geq 0$ as the contribution from a simple expression obtainable in a two-dimensional Euclidian field theory. (Note that we use the Laguerre functions $g_n$ and not the Laguerre polynomials $L_n$ in our approximations). 

We note that the approximation corresponds to the large n-limit  of the harmonic oscillator function, i.e. we are far away from the ground state and consequently ``the motion'' behaves as almost free oscillations.

It is possible inside the same formalism to take the neglected terms in the approximation into account as further contributions in the model. We can use the present results to show that the convergence radius in an expansion of ${\cal R}_N$ around $u=0$ is given by $N^2 m^2$ (just as expected). But further terms in the expansion  are necessary in order to obtain the precise threshold behaviour. We will present such results in future publications.

\section{Concluding Remarks}
\label{Concluding Remarks}

Due to its simple factorisation properties, the Lund Fragmentation Model can as we have shown in here be diagonalised in terms of harmonic oscillator functions. We have, up to now, only treated the $1+1$-dimensional version of the model but we will, in future publications, continue the work into the $3+1$-dimensional real world. 

Transverse momentum is in the fragmentation process of the Lund Model produced via a tunneling mechanism, leading to a gaussian spectrum. In the simplest version of the model, there are no correlations between the transverse momentum generated at one vertex and at the next but the experimental data show, \cite{Eddi}, that such correlations occur at least in the production of the light pions. A mechanism, with strong similarities to the Ornstein-Uhlenbeck process for the velocity distribution of a Brownian motion particle, has been proposed and succesfully applied to the data, \cite{Jim}. 

As this process again is of a gaussian character, we may use very similar methods to diagonalise it, cf. also \cite{ChrisMichael}. There is, however, a small subtlety. If the transverse momentum is firstly generated and afterwards the string-field used to provide longitudinal momentum (as in \textsc{Jetset}) then the transverse mass is used instead of the ordinary mass. This would provide a particular correlation between transverse and longitudinal motion. We will come back in a later publication to a general investigation of the dynamics.
 
Gluon radiation is in the Lund Model treated in terms of internal excitations of the string field and this will lead to a bent string surface, \cite{BA}. The fragmentation of states containing one or more gluons has been introduced into the Lund Model by Sj\"ostrand, \cite{TS2}, using a particular generalisation of the process described above. This process is implemented into \textsc{Jetset} and has been very succesful to describe the experimental data. The method we have introduced above for a flat string surface can be almost directly applied (there is a minor change, that we feel may have some implications for the description of the fastest particles in a gluon jet) to the method introduced by Sj\"ostrand.

Bose-Einstein correlations has been introduced into the Lund Model by interpreting the Lund Area Law as stemming from the square of a quantum mechanical transition matrix element, \cite{WHBA}, \cite{Marcus}. This means that the Area Law for production of two or more identical particles, obtains a weight factor depending upon area differences. One problem in this respect is that with $n$ identical particles the weight factor will obtain contributions from the $n!$ different permutations of the particles. The mathematicians call these problems exponential in $n$ and they are very time-consuming (although we were in \cite{Marcus} able to bring them down considerably). As the area differences are directly expressible in terms of the variables we have discussed above we have some hope to be able to reexpress the full result by means of our formalism.

Finally, we have obtained a new set of tools to study the energy dependence of quantities like $R_N(s)$ and even more interestingly the sum over all the multiplicities $R(s) =\sum_N R_N$. This dependence will necessarily be of the kind $s^a$. This is known from before, \cite{BS}, \cite{BA}, but the power $a$, that corresponds to a ``Regge intercept'' in quark scattering, in accordance with Gribov's Reggeon theory, will in this way be accessible for analytical treatment. Evidently the corresponding power obtained for a ``gluon fragmentation process'' will have some meaning for the soft Pomeron intercept.
 
\section{Acknowledgements}
\label{Acknowledgements}
One of us, Bo Andersson, would like to thank Eddi de Wolfe for suggesting that the Lund Model should be treated by transfer matrix techniques and for providing helpful material.

\end{document}